# Disassembling one-dimensional chains in molybdenum oxides

X. Du[1*], Y. D. Li[1], W. X. Zhao[1], R. Z. Xu[1], K. Y. Zhai[1], Y. L. Chen[2,3,4*], L. X. Yang[1,5,6*]

[1] State Key Laboratory of Low Dimensional Quantum Physics, Department of Physics, Tsinghua University, Beijing 100084, China.

[2] Department of Physics, Clarendon Laboratory, University of Oxford, Parks Road, Oxford OX1 3PU, UK.

[3] School of Physical Science and Technology, ShanghaiTech University and CAS-Shanghai Science Research Center, Shanghai 201210, China.

[4] ShanghaiTech Laboratory for Topological Physics, Shanghai 200031, China.

[5] Frontier Science Center for Quantum Information, Beijing 100084, China.

[6] Collaborative Innovation Center of Quantum Matter, Beijing 100084, China.

e-mails: XD: x-du19@mails.tsinghua.edu.cn; YLC: yulin.chen@physics.ox.ac.uk; LXY: lxyang@tsinghua.edu.cn

**The dimensionality of quantum materials strongly affects their physical properties. Although many emergent phenomena, such as charge-density wave and Luttinger liquid behavior, are well understood in one-dimensional (1D) systems, the generalization to explore them in higher dimensional systems is still a challenging task. In this study, we aim to bridge this gap by systematically investigating the crystal and electronic structures of molybdenum-oxide family compounds, where the contexture of 1D chains facilitates rich emergent properties. While the quasi-1D chains in these materials share general similarities, such as the motifs made up of $MoO_6$ octahedrons, they exhibit vast complexity and remarkable tunability. We disassemble the 1D chains in molybdenum oxides with different dimensions and construct effective models to excellently fit their low-energy electronic structures obtained by *ab initio* calculations. Furthermore, we discuss the implications of such chains on other physical properties of the materials and the practical significance of the effective models. Our work establishes the molybdenum oxides as simple and tunable model systems for studying and manipulating the dimensionality in quantum systems.**

## I. INTRODUCTION

One-dimensional (1D) systems not only exhibit abundant quantum phenomena, such as charge-density wave (CDW) [1], spin-density wave [2], superconductivity [3], magnetism [4], and Luttinger liquid (LL) behavior [5], but also often serve as starting points for modeling complex physics in high-dimensional systems, like the Hubbard model and its variants [6]. Despite their theoretical importance, there are few experimental realizations of exact 1D systems. Instead, 1D physics has mainly been studied in the quasi-1D materials consisting of parallel arrays of 1D chains by neglecting the inter-chain interaction, or artificially constructed nanostructures [7]. Recently, it was noted that 1D physics can also survive in certain high-dimensional systems, including twisted $WTe_2$ [8] and molybdenum oxide $Mo_4O_{11}$ [9].

Molybdenum oxide (MO), with a typical chemical formula $A_xMoO_y$, is a huge materials family [10]. Here $A$ stands for the intercalated metallic atom as primarily an electron donor. The ratio between the numbers of O and Mo atoms, $y$, typically lies between 2 and 3. While crystalline $MoO_3$ ($y = 3$) and $MoO_2$ ($y = 2$) consist purely of $MoO_6$ octahedrons, other members in the family may also contain $MoO_4$ tetrahedrons that can donate electrons. Interestingly, many MOs feature quasi-2D layered structures with hidden quasi-1D chains [11]. The reduction in dimensionality, together with the high tunability of the stoichiometry and crystal structure, makes MO a rich platform for studying and manipulating the effect of dimensionality on the physical properties of real materials.

In the current work, we focus on several low-dimensional MOs with layered structures to study their low-energy electronic structures. We disassemble the crystal structures of MOs into 1D chains, which arrange into either parallel quasi-1D arrays or intercrossed quasi-2D networks. On this basis, we construct low-energy effective models to excellently simulate the electronic structure of these MOs near $E_F$. We propose that, notwithstanding the different dimensionality of MO crystals, their electronic structures can be understood by regarding the materials as assemblies of homologous 1D chains. In particular, we reveal radically different, i.e., quasi-1D and quasi-2D electronic structures of $Li_xMo_6O_{17}$ (x~0.9) and $Mo_4O_{11}$, despite their similar quasi-2D crystal structure, in which the electron filling plays a critical role. Our study provides a unified understanding of the crystal and electronic structures of MO compounds, which sheds light on the understanding and manipulating of the dimensionality in real materials.

The paper is organized as follows. In Sec. II we describe the methodology. Then we construct the effective model for four different materials in Sec. III. Sec. IV presents further discussions on the scientific implications of our results. Finally, the paper concludes in Sec. V.

## II. METHODS

### A. Electronic structure calculation

Electronic structure calculations were performed using density functional theory (DFT) with projected augmented wave method as implemented in the QUANTUM ESPRESSO package [12,13]. The exchange-correlation functional was approximated within the Perdew-Burke-Ernzerhof (PBE) scheme [14]. For each material, the cutoff energy for the plane-wave basis was set to 600 eV and the Monkhorst-Pack $k$-point mesh with a spacing of 0.15 Å$^{-1}$ was employed to get a self-consistent charge density. Subsequently, a non-self-consistent calculation with a finer Monkhorst-Pack $k$-point mesh with a spacing of 0.075 Å$^{-1}$ was performed to evaluate the projected density of states (PDOS) and band energies for the effective model fitting. Surface-projected band structures were computed with the WANNIERTOOLS package [15]. To this end, the tight-binding type Hamiltonians constructed from maximally-localized Wannier functions (MLWF) supplied by the Wannier90 code [16] were exploited. The calculated electronic structures of the studied materials well reproduce experimental results obtained by angle-resolved photoemission spectroscopy [9,17-19].

**B. Effective model**

The effective models are of tight-binding type based on the crystal structure analysis and DFT calculation. The hopping integrals and overall energy shifts were determined by fitting the effective Hamiltonian to the DFT-calculated band energies. We evaluated the effective Hamiltonian numerically for each $k$-point in the uniform $k$-point mesh used in the non-self-consistent calculation described previously. Then we performed least-squares fitting to minimize the difference between the effective-model and the DFT calculated band energies on the $k$-point mesh.

## III. RESULTS

**A. The basic structure of 1D chains in MOs**

The MOs considered here are built with 1D chains of corner-sharing $MoO_6$ octahedrons [Figs. 1(a-c)]. In some MOs, there exist $MoO_4$ tetrahedrons at the edge of the chains as well. The Mo $4d$ orbitals in the octahedrons split into three $t_{2g}$ orbitals and two $e_g$ orbitals. The lowest $t_{2g}$ bands can be filled by the electrons donated by the intercalated metal atoms and/or $MoO_4$ tetrahedrons, which strongly affects the electronic properties of the system (Supplemental Material [20], Fig.

S1).

In general, the corner-sharing octahedrons can form chains via different connection patterns [Figs. 1(d) and (e)]. Both the longitudinal and transverse structures of the chains can vary in different materials. A staggered configuration [Fig. 1(f)] is also possible, which can be viewed as dimerized chains. The widths of the chains are typically 1-2 nm and the contribution to the low-energy electronic structure from the MoO$_6$ octahedrons decays away from the central axis [Fig. 1(c) and Supplemental Material [20], Fig. S1]. In some cases, chains extending along different directions interweave into quasi-2D networks and the three $t_{2g}$ orbitals in a Mo atom can be simultaneously involved.

**B. K$_x$Mo$_6$O$_{17}$**

We first consider K$_x$Mo$_6$O$_{17}$ (x~0.9) in the trigonal space group $P\bar{3}$ (# 147), with lattice parameters $a$ = 13.656 Å, $b$ = $c$ = 5.538 Å, $\alpha$ = 120°, and $\beta$ = $\gamma$ = 90° [21]. The K-intercalated slabs in the crystal structure consist of four octahedron layers encapsulated by two tetrahedron layers [Fig. 2(a)]. Given the $C3$ axis along $a$, each slab can be viewed as woven by equivalent chains running along the $b$, $c$, and $b+c$ directions. In each chain, there are three sets of inequivalent Mo atoms (Mo I, II, and III), as shown in Fig. 2(b).

Figure 2(c) abstracts the slabs as crossed chains with the lattice sites indicated by black dots. For simplification, we consider only the intra-chain hopping $t_n$ between nearest neighbors along C$i$ ($i$ = 1, 2, 3) and the hopping between chains along different directions [C$i$ ↔ C$j$ ($i \neq j$)], $t_{int}$. A 3×3 Hamiltonian can therefore be formulated with the same energy shift $e_s$ for the 3 equivalent chains:

$$H_1 = \begin{pmatrix} -2t_n \cos(\mathbf{k} \cdot \mathbf{r_1}) - e_s & t_{int} & t_{int} \\ t_{int} & -2t_n \cos(\mathbf{k} \cdot \mathbf{r_2}) - e_s & t_{int} \\ t_{int} & t_{int} & -2t_n \cos(\mathbf{k} \cdot \mathbf{r_3}) - e_s \end{pmatrix} \quad (1)$$

with **k** denoting the 2D wave vector in the *b-c* plane and $\mathbf{r_1} = \mathbf{b}$, $\mathbf{r_2} = \mathbf{c}$, $\mathbf{r_3} = \mathbf{b} + \mathbf{c}$. As shown in Fig. 2(d), the model (blue lines) fits remarkably well to the surface-projected DFT calculations (grayscale image) with the best-fitting parameters listed in Table I, despite the sharp contrast between the simplicity of the effective model and the complexity of the crystal structure. Since effectively the two $MoO_4$ tetrahedrons and the ~ 0.9 K atom in a unit cell donate ~2.9 electrons in total, the three low-energy quasi-1D bands are approximately half-filled.

Table I. Fitted parameters of the effective Hamiltonian of $K_xMo_6O_{17}$.

| Parameters | $t_n$ | $t_{int}$ | $e_s$ |
|---|---|---|---|
| Values / meV | 388 | 60 | -18 |

### C. $Mo_4O_{11}$ and its congeners $Mo_9O_{25}$ and $Mo_5O_{14}$

$Mo_4O_{11}$ crystallizes in either monoclinic (η-type) or orthorhombic (γ-type) structure, with two nearly-identical or quasi-mirrored $Mo_8O_{22}$ slabs in a unit cell respectively [22]. The slabs are bridged by $MoO_4$ tetrahedrons (Mo I) with a weak inter-slab coupling. As a result, in the following, we consider only one of the slabs (half of the unit cell).

There are four sets of inequivalent Mo atoms (Mo I-IV) in $Mo_4O_{11}$ [Figs. 3(a)-(c)]. Compared to $K_xMo_6O_{17}$, the reduced symmetry prompts two types of quasi-1D chains along three directions [the chains along $b-c$ (C2) and $b+c$ (C3) are equivalent but different from that along $b$ (C1)] [Fig. 3(c)]. It is worth mentioning that the congeners of $Mo_4O_{11}$ offer the potential to modulate the materials properties via the length of $MoO_6$ octahedron motifs, as exemplified in $Mo_9O_{25}$ and $Mo_5O_{14}$ ($Mo_{10}O_{28}$) [23] (Supplemental Material [20], Note 4).

There are two possibilities to construct effective models for $Mo_4O_{11}$ based on the choice of the structural unit. The first chooses 6 octahedrons in a row in all chains [dashed black and red rectangles in Figs. 3(b) and 3(c) respectively], which leads to a coupled-chain structure similar to $K_xMo_6O_{17}$ (Supplemental Material [20], Fig. S2). The second choice, by contrast, contains clusters of two octahedron rows in C2 and C3 [dashed green rectangle in Fig. 3(c)] as a structural unit, which in turn introduces interacting sublattices with their charge centers marked by yellow stars in Fig. 3(c) (Supplemental Material [20], Note 5). This results in a staggered chain analogous to the 1D Su-Schrieffer-Heeger (SSH) model [Fig. 3(d)], with different hopping integrals between the clusters ($t_{n21}$) and within the same cluster ($t_{n22}$) in the effective Hamiltonian. Apart from $t_{n21}$ and $t_{n22}$, the other hopping integrals are close to those in $K_xMo_6O_{17}$. Consequently, a 5×5 Hamiltonian can be formulated on the basis of (1, 2, 2′, 3, 3′), where distinct energy shifts $e_{s1/2}$ are applied to C1 (site 1) and C2/C3 (site 2/2′/3/3′):

$$H_2 = \begin{pmatrix} -2t_{n1}\cos(\mathbf{k}\cdot\mathbf{r_1}) - e_{s1} & t_{int1} & 0 & t_{int1} & 0 \\ t_{int1} & -e_{s2} & -t_{n21}\exp\left(-i\mathbf{k}\cdot\frac{\mathbf{r_2}}{2}\right) - t_{n22}\exp\left(i\mathbf{k}\cdot\frac{\mathbf{r_2}}{2}\right) & 0 & 0 \\ 0 & -t_{n21}\exp(i\mathbf{k}\cdot\mathbf{r_2}/2) - t_{n22}\exp(-i\mathbf{k}\cdot\mathbf{r_2}/2) & -e_{s2} & 0 & 0 \\ t_{int1} & 0 & 0 & -e_{s2} & -t_{n21}\exp(-i\mathbf{k}\cdot\mathbf{r_3}/2) - t_{n22}\exp(i\mathbf{k}\cdot\mathbf{r_3}/2) \\ 0 & 0 & 0 & -t_{n21}\exp(i\mathbf{k}\cdot\mathbf{r_3}/2) - t_{n22}\exp(-i\mathbf{k}\cdot\mathbf{r_3}/2) & -e_{s2} \end{pmatrix} \quad (2)$$

where $\mathbf{r_1} = \mathbf{b}$, $\mathbf{r_2} = \mathbf{b} - \mathbf{c}$, $\mathbf{r_3} = \mathbf{b} + \mathbf{c}$. Here we assume equal distances between nearby lattice sites along C2/C3. The effective model also agrees nicely with the DFT calculated

(surface-projected) bands in Fig. 3(e), with best-fitting parameters listed in Table II.

Table II. Fitted parameters of the effective Hamiltonian of $Mo_4O_{11}$.

| Parameters | $t_{n1}$ | $t_{n21}$ | $t_{n22}$ | $t_{int1}$ | $t_{int2}$ | $e_{s1}$ | $e_{s2}$ |
|---|---|---|---|---|---|---|---|
| Values / meV | 391 | 503 | 564 | 72 | 52 | -50 | -385 |

For $Mo_4O_{11}$ (also $Mo_9O_{25}$ and $Mo_5O_{14}$), each of the two $MoO_4$ tetrahedrons in a slab (half of the unit cell) donates 2 electrons. Hence, the low-energy bands are occupied with 4 electrons in total. It is worth noting that with the sublattice included, the effective model with staggered hopping integrals introduces a small energy gap above $E_F$, which fits better to the DFT calculated results than the case without the sublattice (See Supplemental Material [20], Fig. S2).

## D. $Li_xMo_6O_{17}$

$Li_xMo_6O_{17}$ (x~0.9) crystallizes into the monoclinic space group $P2_1/m$ (# 11), with lattice parameters $a = 12.75$ Å, $b = 5.524$ Å, $c = 9.491$ Å, $\alpha = 90°$, $\beta = 90.59°$, and $\gamma = 90°$ [24]. There are two chemical formula units in a unit cell [Fig. 4(a)]. Similar to $Mo_4O_{11}$, $Li_xMo_6O_{17}$ features inequivalent chains along $b-c$ ($b+c$) and $b$. But it incorporates two chains related by inversion symmetry along $b$, each highly resembling the chain in $K_xMo_6O_{17}$ [Fig. 4(b)].

Again, considering the effect of sublattices in the chains along $b + c$ or $b - c$ [either in the dashed red or the equivalent green rectangles] fits better to the DFT results [Fig. 4(d), see Supplemental Material [20], Fig. S3]. Following the discussion above, a 6×6 Hamiltonian on the basis of (1, 1′, 2, 2′, 3, 3′) can be formulated as:

$$H_3 = \begin{pmatrix} -2t_{n1}\cos(\mathbf{k}\cdot\mathbf{r}_1)-e_{s1} & 0 & t_{int1} & 0 & -t_{n21}\exp(i\mathbf{k}\cdot\mathbf{r}_2/2)-t_{n22}\exp(-i\mathbf{k}\cdot\mathbf{r}_2/2) & t_{int2} \\ 0 & -2t_{n1}\cos(\mathbf{k}\cdot\mathbf{r}_1)-e_{s1} & 0 & t_{int1} & -e_{s2} & 0 \\ t_{int1} & 0 & -e_{s2} & 0 & t_{int1} & 0 \\ 0 & t_{int1} & -t_{n21}\exp(-i\mathbf{k}\cdot\tfrac{\mathbf{r}_2}{2})-t_{n22}\exp(i\mathbf{k}\cdot\tfrac{\mathbf{r}_2}{2}) & t_{int2} & 0 & t_{int1} \\ -t_{n21}\exp(i\mathbf{k}\cdot\mathbf{r}_2/2)-t_{n22}\exp(-i\mathbf{k}\cdot\mathbf{r}_2/2) & -e_{s2} & t_{int2} & 0 & -e_{s2} & -t_{n21}\exp(i\mathbf{k}\cdot\mathbf{r}_3/2)-t_{n22}\exp(-i\mathbf{k}\cdot\mathbf{r}_3/2) \\ t_{int2} & 0 & 0 & t_{int1} & -t_{n21}\exp(-i\mathbf{k}\cdot\mathbf{r}_3/2)-t_{n22}\exp(i\mathbf{k}\cdot\mathbf{r}_3/2) & -e_{s2} \end{pmatrix} \quad (3)$$

where $\mathbf{r}_1 = \mathbf{b}$, $\mathbf{r}_2 = \mathbf{b} - \mathbf{c}$, and $\mathbf{r}_3 = \mathbf{b} + \mathbf{c}$. Fig. 4(e) plots the fit of the effective model to the DFT calculated (surface-projected) bands with best-fitting parameters listed in Table III.

Table III. Fitted parameters of the sublattice-included effective Hamiltonian of $Li_xMo_6O_{17}$.

| Parameters | $t_{n1}$ | $t_{n21}$ | $t_{n22}$ | $t_{int1}$ | $t_{int2}$ | $e_{s1}$ | $e_{s2}$ |
|---|---|---|---|---|---|---|---|
| Values / meV | 387 | 228 | 447 | 52 | 54 | -10 | -46 |

In $Li_xMo_6O_{17}$, the 4 $MoO_4$ tetrahedrons and ~1.8 Li atoms in a unit cell donate ~5.8 electrons in total (doubled with $K_xMo_6O_{17}$). Consequently, the six low-energy bands are nearly half-filled. The notable gap of several hundreds of milli-eV amid the bands associated with C2/C3 chains suggests a more significant divergence between the two hopping integrals in the staggered chains in $Li_xMo_6O_{17}$. This traces back to the larger "dimerization" of the charge centers in $Li_xMo_6O_{17}$ than in $Mo_4O_{11}$ [yellow stars in Figs. 4(c) and 3(c)], which is reminiscent of the Peierls transition (Supplemental Material [20], Note 5).

### E. $K_{0.3}MoO_3$

$K_{0.3}MoO_3$ crystallizes into the monoclinic space group $C2/m$ (# 12), with lattice parameters

$a$ = 18.162 Å, $b$ = 7.554 Å, $c$ = 9.816 Å, $\alpha$ = 90°, $\beta$ = 90.59°, $\gamma$ = 90° [25] and contains 20 (10) chemical formula units in a conventional (primitive) unit cell. The crystal structure of $K_{0.3}MoO_3$ differs from the MOs presented above [Fig. 5(a)]. First, there is solely one extending direction of the quasi-1D chains in the crystal, in contrast to the intercrossed quasi-1D chains introduced before. The chains in $K_{0.3}MoO_3$ arrange parallel to each other, constituting slabs on the $b$-$d$ plane ($\mathbf{d} = \mathbf{a} + 2\mathbf{c}$). Second, there is no $MoO_4$ tetrahedron, and conducting carriers are donated by K atoms only. The structural unit in $K_{0.3}MoO_3$ appears as a double-chain as shown in Fig. 5(b). Every structural unit connects with two neighbors through corner-sharing Mo III octahedrons, generating staircases in the $b$-$d$ plane [Fig. 5(a)], with each step made up of two chains [C1 and C1′, see Figs. 5 (a, c, d)]. The hopping integrals to be considered include the hopping within each chain ($t_n$), and hopping between chains within the same step ($t_{int1}$) and in nearby steps ($t_{int2}$) of the staircase. A 2×2 Hamiltonian can therefore be formulated as:

$$H_4 = \begin{pmatrix} -2t_n \cos(\mathbf{k} \cdot \mathbf{r}_1) - e_s & -t_{int1}[\exp(-i\mathbf{k} \cdot \boldsymbol{\delta}_1) + \exp(-i\mathbf{k} \cdot \boldsymbol{\delta}_2)] - t_{int2} \exp(-i\mathbf{k} \cdot \boldsymbol{\delta}_3) \\ -t_{int1}[\exp(i\mathbf{k} \cdot \boldsymbol{\delta}_1) + \exp(i\mathbf{k} \cdot \boldsymbol{\delta}_2)] - t_{int2} \exp(i\mathbf{k} \cdot \boldsymbol{\delta}_3) & -2t_n \cos(\mathbf{k} \cdot \mathbf{r}_1) - e_s \end{pmatrix} \quad (4)$$

with $\mathbf{r}_1 = \mathbf{b}$, $\boldsymbol{\delta}_1 = -\frac{1}{2}(\mathbf{a} + \mathbf{b}) + \boldsymbol{\delta}_0$, $\boldsymbol{\delta}_2 = -\frac{1}{2}(\mathbf{a} - \mathbf{b}) + \boldsymbol{\delta}_0$, $\boldsymbol{\delta}_3 = \mathbf{c} + \boldsymbol{\delta}_0$, where $\boldsymbol{\delta}_0$ is the relative position between sites on C1 and C1′ in the same structural unit (whose absolute value does not affect the resultant band structure). The fitting to the DFT calculated bands along high-symmetry directions is plotted in Fig. 5(e) with best-fitting parameters listed in Table IV.

Considering that the lattice sites in C1 and C1′ within the same step [Fig. 5(c)] constitute a triangular ladder and the hopping integrals $t_n$ and $t_{int1}$ are nearly equal, the effective model can be further simplified to a two-leg triangular ladder model, which provides a rare opportunity to theoretically explore the emergent properties in a complex real material, such as the Luttinger liquid behavior and electron-phonon interaction [18,26,27] (See Supplemental Material [20],

Fig. S4).

Table IV. Fitted parameters of the effective Hamiltonian of $K_{0.3}MoO_3$.

| Parameters | $t_n$ | $t_{int1}$ | $t_{int2}$ | $e_s$ |
|---|---|---|---|---|
| Values / meV | 204 | 215 | 32 | 328 |

The three K atoms within a primitive unit cell lead to a 3/4-filling of the two lowest bands. If there is a lower filling factor, for example, 1/4-filling, only 1 band would intersect $E_F$, and then the low-energy band could be alternatively described by a single uniform chain, which is made up of the entire step in the staircase consisting of one C1 chain and one C1′ chain [Fig. 5(c, d)], thereby emulating the situation depicted in Fig. 1(e) (also see Supplemental Material [20], Fig. S1).

**F. Other MO-related compounds**

In addition to the materials introduced above, the vast MO family contains many other congeners and isomers with similar chain-like structures and can therefore be analyzed likewise. Supplemental Note 7 [20] presents more details about the isomers. Moreover, there exists a larger family of tungsten bronzes with chain structures similar to $Mo_4O_{11}$, including monophosphate tungsten bronze (MPTB) and diphosphate tungsten bronze (DPTB) [28]. They appear with general chemical formulas of $A_x(PO_2)_4(WO_3)_{p+q}$ and $A_x(P_2O_4)_2(WO_3)_{p+q}$ respectively, where $A$ is the intercalated metal atom. The variation of integers p and q, together with the alternation of $A$ and x, greatly enriches the abundance of the tungsten bronze family.

In MPTBs and DPTBs, the structure of coupled-1D chains is highly similar to that of MOs, with $MoO_6$ octahedrons ($MoO_4$ tetrahedrons) replaced by $WO_6$ octahedrons ($PO_4$ tetrahedrons or $P_2O_7$ double-tetrahedrons). Such a replacement lowers the filling of the low-energy quasi-1D bands, due to the shortage of valence electrons of P compared with Mo. A notable characteristic of the tungsten bronzes is the larger spin-orbit coupling (SOC) strength of W atoms, which may be attractive for the establishment of SOC-induced spin transport effects [29].

**IV. Discussion**

**A. Uniqueness of Mo-oxides**

Many MOs are derived from $MoO_3$ by replacing some $MoO_6$ octahedrons with $MoO_4$ tetrahedrons during the reduction process [30]. The same viewpoint applies to MPTBs / DPTBs, where $PO_4$ / $P_2O_7$ tetrahedrons substitute for some of the $WO_6$ octahedrons in $WO_3$. Such a replacement can be more energetically favorable than oxygen site vacancies [30,31]. As a result, the 3D network of $ReO_3$-type octahedrons transforms into disconnected quasi-2D slabs. The crystal field splitting of Mo $4d$ orbitals into three directional $t_{2g}$ components, together with the corresponding charge configuration, induces the orbital-selective quasi-1D electronic structure, which rationalizes the disassembly of the quasi-2D slabs into 1D chains.

On the other hand, the 6-coordinated Mo-O octahedron (with an effective coordination number of 3 due to the corner-sharing connection) and the 6 valence electrons of Mo atoms result in a closed shell with $E_F$ lying between the O $p$ band continuum and the Mo $d$ band continuum. Therefore, the lowest quasi-1D bands by Mo $t_{2g}$ orbitals are occupied by electrons donated by

MoO$_4$ tetrahedrons and intercalated metal atoms. For valence electron number different from 6, the quasi-1D orbitals, if exist, are either empty or fully occupied, making the electronic structure practically 2D / 3D. Therefore, with more delocalized 4d or 5d orbitals, Mo and W atoms are ideal for the formation of quasi-1D electronic structures in the quasi-2D crystals.

**B. Comparison of the quasi-1D chains in different materials**

Although the detailed structures and chemical environments of the quasi-1D chains are diverse, they manifest their structural rigidity with similar bond length and corner-sharing manner along the chain direction in different MOs, as summarized in Fig. 1. Consequently, the intra-chain hopping integrals coincide across different chains in different materials and dominate over other hopping integrals in the electronic structures.

On the other hand, the electronic structure of MOs can be strongly modulated by the fine structure of the chains. The "dimerization" of the staggered chains (C2/C3 in Mo$_4$O$_{11}$ and Li$_{0.9}$Mo$_6$O$_{17}$) induces band gaps in Mo$_4$O$_{11}$, which are further enhanced in Li$_{0.9}$Mo$_6$O$_{17}$ (Supplemental Material [20], Note 5). In Mo$_4$O$_{11}$, the energy bands of the staggered chains are close to 3/8-filling, and therefore the gaps in the bands locate far away from $E_F$ by a distance of 200 meV or more. Conversely, in Li$_x$Mo$_6$O$_{17}$, the half-filling of the staggered chain bands forces $E_F$ to cut amid the large gap of > 200 meV. The low-energy electrons that dominate the physical properties of the material mainly occupy the nearly-degenerate quasi-1D double bands that originate from the C1 and C1′ chains. Therefore, despite the quasi-2D crystal structure, the dimensionality of the low-energy electronic structure is primarily quasi-1D, in drastic contrast to Mo$_4$O$_{11}$. In other words, the presence of staggered chains and hence the band gaps enable the control of the electron dimensionality by electron doping. The chemical potential can be tuned by the content of intercalated metal atoms [28], while the congeners in the $Mo_4O_{11}$ family exemplify the independent tuning of the energy gap (Supplemental Material [20], Note 4).

**C. Implications of the quasi-1D structure**

Thus far, we concentrated primarily on the crystal and band structures of the materials. However, many other physical phenomena have been experimentally explored, which warrant a detailed examination since they are rooted in intrinsic quasi-1D physics.

A significant hallmark to consider in the quasi-1D realm is LL physics. The LL theory is initially designed for 1D systems, where the reduced phase space relative to higher dimensional systems invalidates the Fermi liquid theory [5]. The LL behavior in $Li_xMo_6O_{17}$ has long been experimentally confirmed [19], as anticipated from its nearly-perfect 1D electronic structure near $E_F$ despite the quasi-2D crystal structure as discussed earlier. The inherent quasi-1D material, $K_{0.3}MoO_3$, is also a material with a LL normal state [18]. Interestingly, similar LL physics has been experimentally demonstrated in the quasi-2D $Mo_4O_{11}$ [9], where the orthogonality between the Mo $t_{2g}$ atomic orbitals plays a crucial role.

Another noteworthy phenomenon of quasi-1D materials is the transition between LL and CDW. Among the 4 materials that have been detailed, 3 of them, including $K_xMo_6O_{17}$, $Mo_4O_{11}$, and $K_{0.3}MoO_3$, display CDW transitions at low temperatures. All of them feature periodic lattice distortion and gap opening in the electronic structure, with corresponding CDW wave vectors agreeing perfectly with the predictions by extracting the filling factors from the effective models, as detailed in Supplemental Material [20], Note 8. By contrast, $Li_xMo_6O_{17}$ exhibits an undefined transition at 26 K, followed by a superconductivity transition at 1.8 K [19].

Considering the absence of well-defined quasiparticles in the LL normal phase in the 4 materials, their phase transitions require special treatment. Thus, the MO family provides a rare platform to explore the LL-originated phase transitions in real materials.

**D. Advanced modeling based on the simple effective models**

As elucidated above, the LL physics in MOs can be qualitatively depicted by the 1D electronic structures. For the extraction of interacting parameters of the materials, quantitative models are essential. In fact, there have been extensive theoretical investigations starting from the double-chain structure (C1/1′) of $Li_xMo_6O_{17}$ to study the LL physics [32] and superconductivity transition [33]. We here emphasize that the presence of other chains (C2 and C3) can also impact on the physical properties, for example, a contribution to the downfolding of the dimensional crossover energy scale [34]. For the LL physics in $K_{0.3}MoO_3$, $Mo_4O_{11}$, and $K_xMo_6O_{17}$, a quantitative LL theory is still lacking yet, especially for a 2D LL theory in realistic materials.

The CDW transitions in $K_xMo_6O_{17}$, $Mo_4O_{11}$, and $K_{0.3}MoO_3$ can, to some extent, be explained with the quasi-1D chain structures under the Peierls scenario [1], such as the close relation between the CDW wave vector and electron filling factor. Therefore, for a quantitative CDW theory of these materials, it is a good choice to start from the quasi-1D chains as described in the current work, and incorporate the interplay between various degrees of freedom to enhance accuracy and tackle unsolved problems. An extra yet vital factor to incorporate is LL physics, which can in principle leads to distinct theories with the ones starting from the Fermi liquid normal state. The phase transitions in $Li_xMo_6O_{17}$ may also be understood in the same manner.

The current effective Hamiltonians are well-suited for analytical or calculation-demanding theories, but may not be accurate enough for certain numerical calculations due to the trade-off between model size and precision. As a solution, higher-order hopping terms can be added to the Hamiltonian, such as next-nearest neighbor hopping, hopping between chains along the same directions, and hopping beyond the same unit cell. In Supplemental Material [20], Note 9, we formulate such refined effective models for the 4 materials respectively.

Alternatively, one can build tight-binding models based on Wannier functions [35] that are interfaced with DFT codes, with only the lowest-energy bands included. By this means, the DFT band structure can be perfectly reproduced, at the expense of tremendous parameters in the resultant model. The size of the Wannier-based model can be limitedly controlled, at least in principle, by the fineness of the uniform $k$-point mesh in sampling the Brillouin zone.

**V. Conclusion**

The current work examines the crystal and electronic structures of several low-dimensional MOs in detail. Disassembling their crystal structures yields 1D chains of Mo-O octahedrons, based on which effective Hamiltonians of the weakly-interacting-chain models are constructed, which fit well to the real band structures. Our work not only presents a rare materials platform with rich tunability of the structural and electronic dimensionality but also provides a simple model for the theoretical exploration of emergent properties of complex oxides, such as the LL behavior and its competition/interaction with electronic orders.

## References

[1]  G. Grüner, The dynamics of charge-density waves, Rev. Mod. Phys. **60**, 1129 (1988).

[2]  G. Grüner, The dynamics of spin-density waves, Rev. Mod. Phys. **66**, 1 (1994).

[3]  K. Y. Arutyunov, D. S. Golubev, and A. D. Zaikin, Superconductivity in one dimension, Phys. Rep. **464**, 1 (2008).

[4]  H.-J. Mikeska and A. K. Kolezhuk, *Quantum Magnetism* 2004), Lect. Notes Phys.

[5]  J. Voit, One-dimensional Fermi liquids, Rep. Prog. Phys. **58**, 977 (1995).

[6]  D. P. Arovas, E. Berg, S. A. Kivelson, and S. Raghu, The Hubbard Model, Annu. Rev. Condens. Matter Phys. **13**, 239 (2022).

[7]  V. V. Deshpande, M. Bockrath, L. I. Glazman, and A. Yacoby, Electron liquids and solids in one dimension, Nature **464**, 209 (2010).

[8]  P. Wang *et al.*, One-dimensional Luttinger liquids in a two-dimensional moiré lattice, Nature **605**, 57 (2022).

[9]  X. Du *et al.*, Crossed Luttinger liquid hidden in a quasi-two-dimensional material, Nat. Phys. **19**, 40 (2022).

[10] M. Greenblatt, Molybdenum oxide bronzes with quasi-low-dimensional properties, Chem. Rev. **88**, 31 (1988).

[11] E. Canadell and M. H. Whangbo, Conceptual aspects of structure-property correlations and electronic instabilities, with applications to low-dimensional transition-metal oxides, Chem. Rev. **91**, 965 (1991).

[12] P. Giannozzi *et al.*, QUANTUM ESPRESSO: a modular and open-source software project for quantum simulations of materials, J. Phys.: Condens. Matter **21**, 395502 (2009).

[13] P. Giannozzi *et al.*, Advanced capabilities for materials modelling with Quantum ESPRESSO, J. Phys.: Condens. Matter **29**, 465901 (2017).

[14] J. P. Perdew, K. Burke, and M. Ernzerhof, Generalized gradient approximation made simple, Phys. Rev. Lett. **77**, 3865 (1996).

[15] Q. Wu, S. Zhang, H.-F. Song, M. Troyer, and A. A. Soluyanov, WannierTools: An open-source software package for novel topological materials, Comput. Phys. Commun. **224**, 405 (2018).

[16] G. Pizzi *et al.*, Wannier90 as a community code: new features and applications, J. Phys.: Condens. Matter **32**, 165902 (2020).

[17] D. Mou *et al.*, Discovery of an Unconventional Charge Density Wave at the Surface of $K_{0.9}Mo_6O_{17}$, Phys. Rev. Lett. **116**, 196401 (2016).

[18] L. Kang *et al.*, Band-selective Holstein polaron in Luttinger liquid material $A_{0.3}MoO_3$ ($A$ = K, Rb), Nat. Commun. **12**, 6183 (2021).

[19] L. Dudy, J. D. Denlinger, J. W. Allen, F. Wang, J. He, D. Hitchcock, A. Sekiyama, and S. Suga, Photoemission spectroscopy and the unusually robust one-dimensional physics of lithium purple bronze, J. Phys.: Condens. Matter **25**, 014007 (2013).

[20] See Supplemental Material at [URL] for the calculation of PDOS of separated materials, simplified effective model for $Mo_4O_{11}$, $Li_xMo_6O_{17}$, and $K_{0.3}MoO_3$, discussion of congeners of $Mo_4O_{11}$ and additional molybdenum oxide compounds, details about dimerized chains in $Mo_4O_{11}$ and $Li_xMo_6O_{17}$, discussion about CDW properties in the MO compounds and refined effective models.

[21] H. Vincent, M. Ghedira, J. Marcus, J. Mercier, and C. Schlenker, Structure cristalline d'un

conducteur métallique bidimensionnel: Le bronze violet de potassium et molybdene $K_{0.9}Mo_6O_{17}$, J. Solid State Chem. **47**, 113 (1983).

[22] L. Kihlborg, Crystal structure studies on monoclinic and orthorhombic $Mo_4O_{11}$, Arkiv for Kemi **21**, 365 (1963).

[23] F. Portemer, M. Sundberg, L. Kihlborg, and M. Figlarz, Homologues of $Mo_4O_{11}$ (mon) in the Mo-W-O System Prepared by Soft Chemistry, J. Solid State Chem. **103**, 403 (1993).

[24] M. S. da Luz, J. J. Neumeier, C. A. M. dos Santos, B. D. White, H. J. I. Filho, J. B. Leão, and Q. Huang, Neutron diffraction study of quasi-one-dimensional lithium purple bronze: Possible mechanism for dimensional crossover, Phys. Rev. B **84**, 014108 (2011).

[25] W. J. Schutte and J. L. de Boer, The incommensurately modulated structures of the blue bronzes $K_{0.3}MoO_3$ and $Rb_{0.3}MoO_3$, Acta Crystallogr., Sect. B: Struct. Sci **49**, 579 (1993).

[26] L. Perfetti, S. Mitrovic, G. Margaritondo, M. Grioni, L. Forró, L. Degiorgi, and H. Höchst, Mobile small polarons and the Peierls transition in the quasi-one-dimensional conductor $K_{0.3}MoO_3$, Phys. Rev. B **66**, 075107 (2002).

[27] D. Mou, R. M. Konik, A. M. Tsvelik, I. Zaliznyak, and X. Zhou, Charge-density wave and one-dimensional electronic spectra in blue bronze: Incoherent solitons and spin-charge separation, Phys. Rev. B **89**, 201116 (2014).

[28] P. Roussel, O. Pérez, and P. Labbé, Phosphate tungsten bronze series: crystallographic and structural properties of low-dimensional conductors, Acta Crystallogr., Sect. B: Struct. Sci **57**, 603 (2001).

[29] C. H. L. Quay, T. L. Hughes, J. A. Sulpizio, L. N. Pfeiffer, K. W. Baldwin, K. W. West, D. Goldhaber-Gordon, and R. de Picciotto, Observation of a one-dimensional spin–orbit gap in a

quantum wire, Nat. Phys. **6**, 336 (2010).

[30] K. Inzani, M. Nematollahi, F. Vullum-Bruer, T. Grande, T. W. Reenaas, and S. M. Selbach, Electronic properties of reduced molybdenum oxides, Physical Chemistry Chemical Physics **19**, 9232 (2017).

[31] Y.-J. Lee, T. Lee, and A. Soon, Phase Stability Diagrams of Group 6 Magnéli Oxides and Their Implications for Photon-Assisted Applications, Chem. Mater. **31**, 4282 (2019).

[32] P. Chudzinski, T. Jarlborg, and T. Giamarchi, Luttinger-liquid theory of purple bronze $Li_{0.9}Mo_6O_{17}$ in the charge regime, Phys. Rev. B **86**, 075147 (2012).

[33] W. Cho, C. Platt, R. H. McKenzie, and S. Raghu, Spin-triplet superconductivity in a weak-coupling Hubbard model for the quasi-one-dimensional compound $Li_{0.9}Mo_6O_{17}$, Phys. Rev. B **92**, 134514 (2015).

[34] T. Giamarchi, Theoretical framework for quasi-one dimensional systems, Chem. Rev. **104**, 5037 (2004).

[35] N. Marzari, A. A. Mostofi, J. R. Yates, I. Souza, and D. Vanderbilt, Maximally localized Wannier functions: Theory and applications, Rev. Mod. Phys. **84**, 1419 (2012).

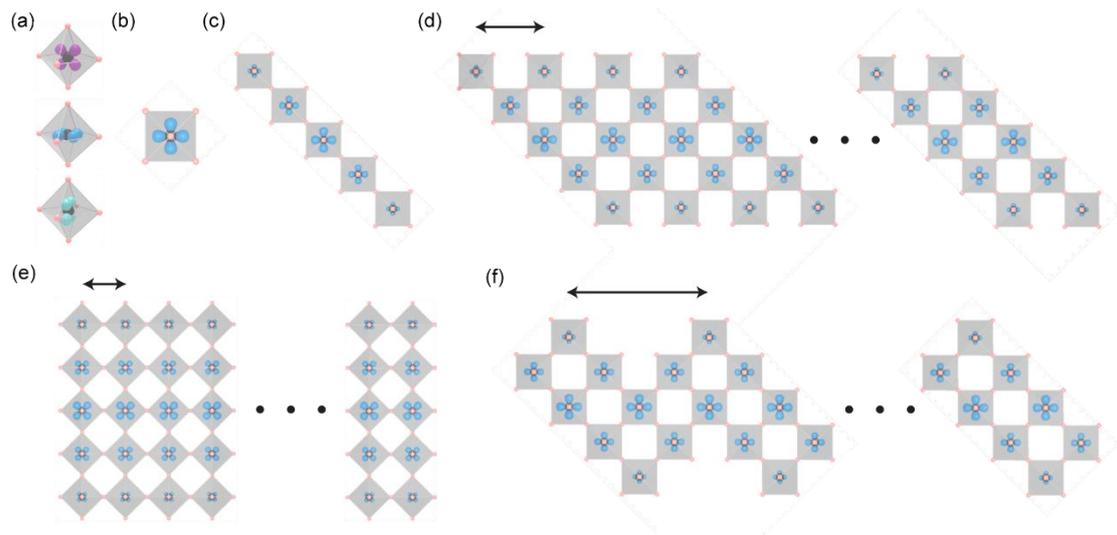

FIG. 1. (a) The three $t_{2g}$ orbitals in a $MoO_6$ octahedron marked with different colors. (b) One of the $t_{2g}$ orbitals viewing from its normal direction. (c) The motif of the quasi-1D Mo-O octahedron chain. The electron density decays away from the chain center, as indicated by the size of the $t_{2g}$ orbital. (d, e) Quasi-1D chains constructed by repeating the motifs, connected in different ways. (f) A quasi-1D chain constructed by repeating an asymmetric double-row motif. The directions of arrows in (d-f) represent the extending direction of quasi-1D chains and the lengths of the arrows represent the unit lengths along the chain.

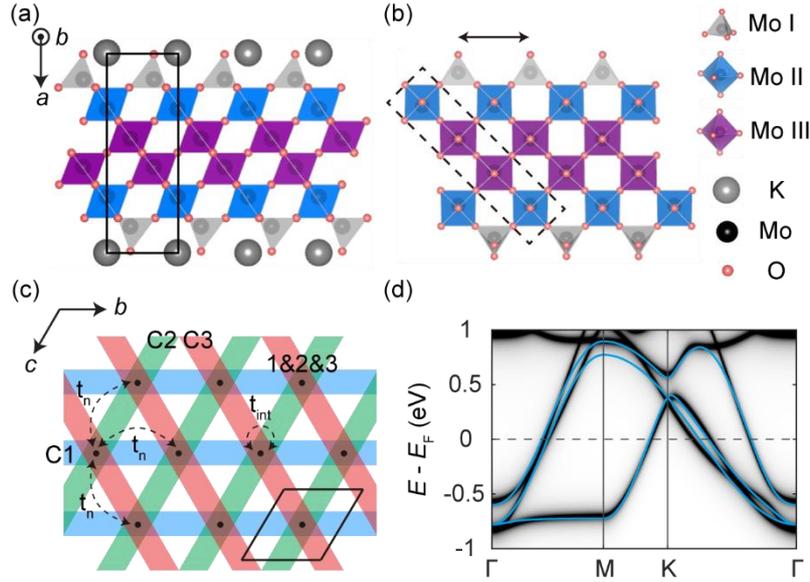

FIG. 2. (a) Crystal structure of $K_xMo_6O_{17}$ viewing along the $b$ direction, with a section of the unit cell indicated with the parallelogram. (b) Quasi-1D chain-like structure viewing along its normal direction. The direction and length of the arrow indicate the extending direction of the chain ($b$, $c$, or $b+c$) and the unit length along the chain, respectively. The dashed rectangle includes a structural unit of the quasi-1D chain that corresponds to Fig. 1(d). (c) Schematic of the effective model within the $b$-$c$ plane. There are three equivalent orbitals that extend along three equivalent directions in a unit cell (black rhombus), represented by 1-3. $t_n$ and $t_{int}$ are the inequivalent hopping integrals included in the effective model (see the main text). (d) Comparison of the band structure along high-symmetry $k$-paths between the surface-projected *ab initio* calculation (grayscale image) and effective model calculation (blue lines).

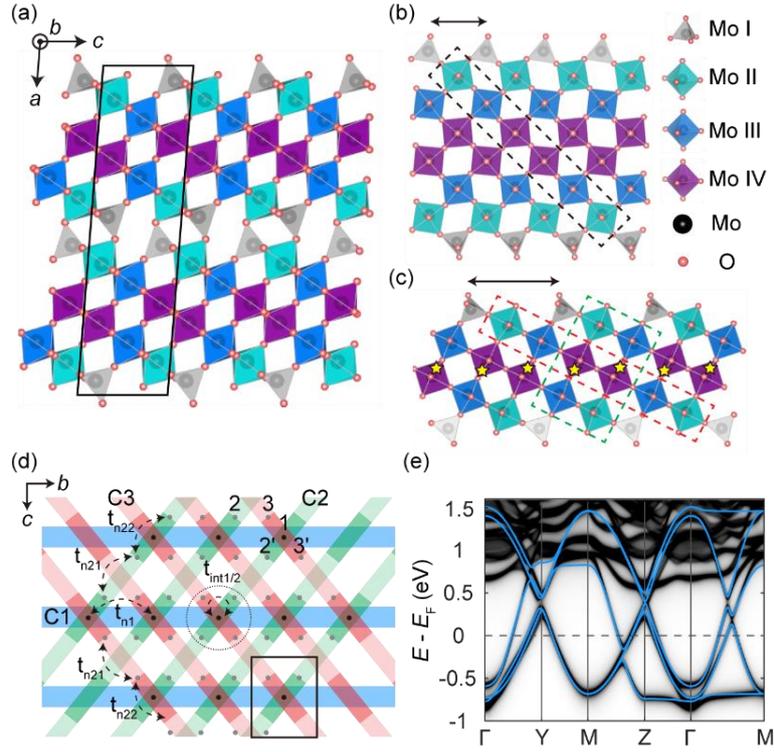

FIG. 3. (a) Crystal structure of $Mo_4O_{11}$ viewing along the *b* direction, with a section of the unit cell indicated with the parallelogram. (b, c) Quasi-1D chain-like structures viewing their normal directions. The directions and lengths of the arrows indicate the extending directions of the chains [*b* for (b) and *b+c* / *b-c* for (c)] and the unit lengths along the chains, respectively. Each of the dashed black rectangle in (b) and red / green rectangle in (c) includes a structural unit of the quasi-1D chain. The yellow stars indicate the rough positions of the charge centers of the 2 octahedron rows in the dashed green rectangle and equivalent positions. (d) Schematic of the effective model with sublattice included within the *b-c* plane. There are five orbitals in a unit cell (black rectangle) that extend along three directions, represented by 1, 2/2', and 3/3'. $t_{n1}$, $t_{n21}$, $t_{n22}$, $t_{int1}$, and $t_{int2}$ are the inequivalent hopping integrals included in the effective model (see the main text). (e) Comparison of the band structure along high-symmetry *k*-paths between the surface-projected *ab initio* calculation (grayscale image) and the calculation with the effective model (blue lines).

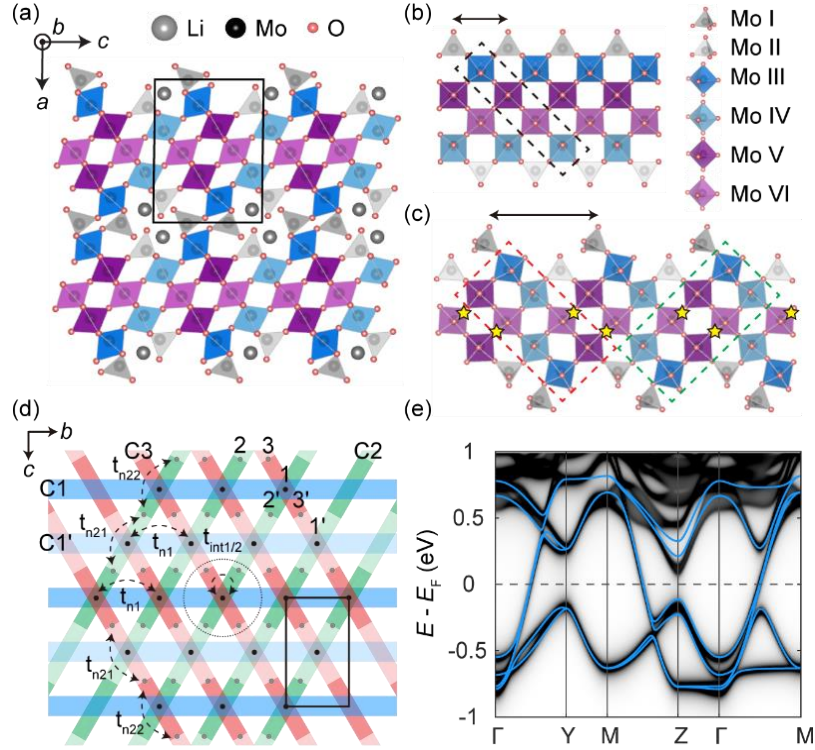

FIG. 4. (a) Crystal structure of Li$_x$Mo$_6$O$_{17}$ viewing along the $b$ direction, with a section of the unit cell indicated with the black parallelogram. (b, c) Quasi-1D chain-like structures viewing their normal directions. The directions and lengths of the arrows indicate the extending directions of the chains [$b$ for (b) and $b+c$ / $b-c$ for (c)] and the unit lengths along the chains, respectively. Each of the dashed black rectangle in (b) and dashed red / green rectangle in (c) includes a structural unit of the quasi-1D chain. The yellow stars indicate the rough positions of the charge centers of the 2 octahedron rows in the dashed green rectangle and equivalent positions. (d) Schematic of the effective model with sublattice included within the $b$-$c$ plane. There are six orbitals in a unit cell (black rectangle) that extend along three directions, represented by 1/1', 2/2', and 3/3'. $t_{n1}$, $t_{n21}$, $t_{n22}$, $t_{int1}$, and $t_{int2}$ are the inequivalent hopping integrals included in the effective model (see the main text). (e) Comparison of the band structure along high-symmetry $k$-paths between the surface-projected *ab initio* calculation (grayscale image) and the calculation with the effective model (blue lines).

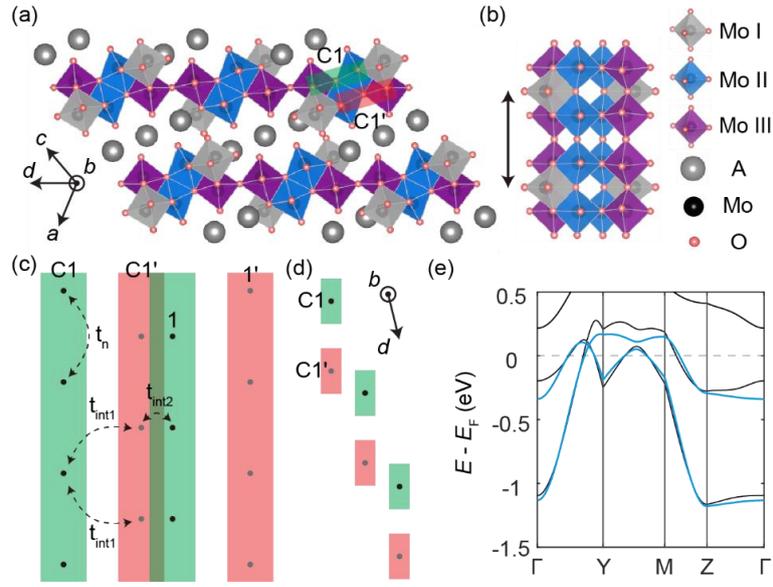

FIG. 5. (a) Crystal structure of $K_{0.3}MoO_3$ viewing along the *b* direction. The axis labels shown are for the conventional unit cell. The direction along which the staircases of quasi-1D chains ascend and descend is marked with **d**=**a**+2**c**. (b) Quasi-1D chain-like structure viewing its normal direction. The direction and length of the black arrow denote the extending direction of the chain (*b*) and the unit length along the chain, respectively. (c, d) Schematic of the effective model, viewing from the normal direction of the chain (c) and the *b* direction (d). There are two equivalent orbitals that extend along the same direction in a unit cell, represented by 1/1'. $t_n$, $t_{int1}$, and $t_{int2}$ are the inequivalent hopping integrals included in the effective model (see the main text). (e) Comparison of the band structure along high-symmetry *k*-paths between the *ab initio* calculation (black lines) and calculation with the effective model (blue lines).